\documentclass[12pt]{iopart}

\usepackage{iopams}

\usepackage{cite}
\usepackage{graphicx}
\usepackage{textcomp}
\usepackage{xcolor}

\usepackage[ruled,vlined]{algorithm2e}

\usepackage[english]{babel}

\usepackage{pstool}                    
\usepackage[dvips]{rotating,graphicx}  
\usepackage{float}                     
\usepackage{examplep}

\usepackage{hyperref}

\usepackage{enumitem}  
\usepackage{latexsym}
\usepackage[english]{babel}
\usepackage{csquotes}

\usepackage{siunitx}                        
\usepackage{bm}  

\usepackage{multirow}       
\usepackage{booktabs}       
\usepackage{makecell}       
\usepackage{tikz}

\usepackage{ragged2e}
\usepackage{xcolor} 
\usepackage{graphicx} 
\usepackage{cite}
\usepackage[nameinlink,capitalize]{cleveref}  

\usepackage{float}

\begin{document}

\title[Validating Large-Scale Quantum Machine Learning using cuTN-QSVM]{Validating Large-Scale Quantum Machine Learning: Efficient Simulation of Quantum Support Vector Machines Using Tensor Networks}

\author{Kuan-Cheng Chen\textsuperscript{1,2*}, Tai-Yue Li\textsuperscript{3*}, Yun-Yuan Wang\textsuperscript{4*},\\ Simon See\textsuperscript{5}, Chun-Chieh Wang\textsuperscript{6}, Robert Wille\textsuperscript{7},\\ Nan-Yow Chen\textsuperscript{8}, An-Cheng Yang\textsuperscript{8}, and Chun-Yu Lin\textsuperscript{8}}

\address{\textsuperscript{1}Department of Electrical and Electronic Engineering, Imperial College London, London, United Kingdom}
\address{\textsuperscript{2}QuEST, Imperial College London, London, United Kingdom}
\address{\textsuperscript{3}National Synchrotron Radiation Research Center, Hsinchu, Taiwan}
\address{\textsuperscript{4}NVIDIA AI Technology Center, NVIDIA Corp., Taipei, Taiwan}
\address{\textsuperscript{5}NVIDIA AI Technology Center, NVIDIA Corp., Santa Clara, CA, USA}
\address{\textsuperscript{6}National Synchrotron Radiation Research Center, Hsinchu, Taiwan}
\address{\textsuperscript{7}Chair of Design Automation, Technical University of Munich, Munich, Germany}
\address{\textsuperscript{8}National Center for HPC, Narlabs, Hsinchu, Taiwan}

\begin{indented}
\item[] * The first three authors contributed equally to this work.
\end{indented}
\ead{kuan-cheng.chen17@imperial.ac.uk}
\vspace{10pt}

\vspace{10pt}
\begin{indented}
\item[]December 2024
\end{indented}

\begin{abstract}
We present an efficient tensor-network-based approach for simulating large-scale quantum circuits exemplified by Quantum Support Vector Machines (QSVMs). Experimentally, leveraging the cuTensorNet library on multiple GPUs, our method effectively reduces the exponential runtime growth to near-quadratic scaling with respect to the number of qubits in practical scenarios. Traditional state-vector simulations become computationally infeasible beyond approximately 50 qubits; in contrast, our simulator successfully handles QSVMs with up to 784 qubits, executing simulations within seconds on a single high-performance GPU. Furthermore, utilizing the Message Passing Interface (MPI) for multi-GPU environments, our method demonstrates strong linear scalability, effectively decreasing computation time as dataset sizes increase. We validate our framework using the MNIST and Fashion MNIST datasets, achieving successful multiclass classification and highlighting the potential of QSVMs for high-dimensional data analysis. By integrating tensor-network techniques with advanced high-performance computing resources, this work demonstrates both the feasibility and scalability of simulating large-qubit quantum machine learning models, providing a valuable validation tool within the emerging Quantum–HPC ecosystem.
\end{abstract}
\clearpage

\section{Introduction}
In the rapidly evolving landscape of artificial intelligence (AI), machine learning algorithms stand out as pivotal components driving advancements across a multitude of domains\cite{jordan2015machine}. These algorithms, distinguished into supervised and unsupervised learning paradigms, harness the power of data to uncover patterns or make predictions\cite{mackay2003information}. Supervised learning, in particular, leverages pre-labeled data to train models, with the Support Vector Machine (SVM) being a cornerstone technique in this category\cite{hearst1998support}. SVMs excel in classifying data into distinct categories by finding an optimal hyperplane in either the original or a higher-dimensional feature space\cite{ghaddar2018high}. However, the computational demands of SVMs, especially in the context of large-scale ``big data" applications\cite{gaye2021improvement}, pose significant challenges in terms of both computational resources and execution time.

Enter the realm of quantum computing, a burgeoning field offering profound computational speedups over classical approaches for certain problem types. Among these, Quantum Support Vector Machines (QSVMs) emerge as a promising quantum-enhanced technique for machine learning\cite{rebentrost2014quantum,wang2021towards,li2022BIBE}, capable of drastically reducing the computational resources required for SVMs. Leveraging quantum algorithms, QSVMs achieve exponential speedups in both training and classification tasks by performing calculations in parallel and employing quantum-specific optimizations\cite{rebentrost2014quantum,li2015experimental,ding2021quantum,chen2023quantum}.

However, in the current Noisy Intermediate-Scale Quantum (NISQ) era\cite{preskill2018quantum}, the practical utility of quantum computers is significantly constrained by their availability and imperfect technological state. Challenges such as the fidelity of qubits, the error rates of two-qubit gates, and the limited number of available qubits present substantial hurdles\cite{kjaergaard2020superconducting,bruzewicz2019trapped,evered2023high}. Despite the advent of several methodologies aimed at enhancing qubit fidelity—such as Quantum Error Mitigation\cite{cai2023quantum, chen2023short} and Dynamical Decoupling\cite{souza2011robust}—these limitations persist, impeding the realization of quantum advantage on quantum computing platforms in the current NISQ era\cite{lau2022nisq,chen2023complexity}. Consequently, the design and validation of quantum-inspired algorithms, or hybrid classical-quantum algorithms, are predominantly conducted through high-performance classical simulations\cite{chen2023quantum,chen2024quantum}. Furthermore, quantum circuit simulators have shown considerable success in the near-term verification of quantum algorithms on small qubit systems\cite{lykov2023fast,chen201864}. 

\begin{figure}[t]
    \centering
    \includegraphics[width=0.75\textwidth]{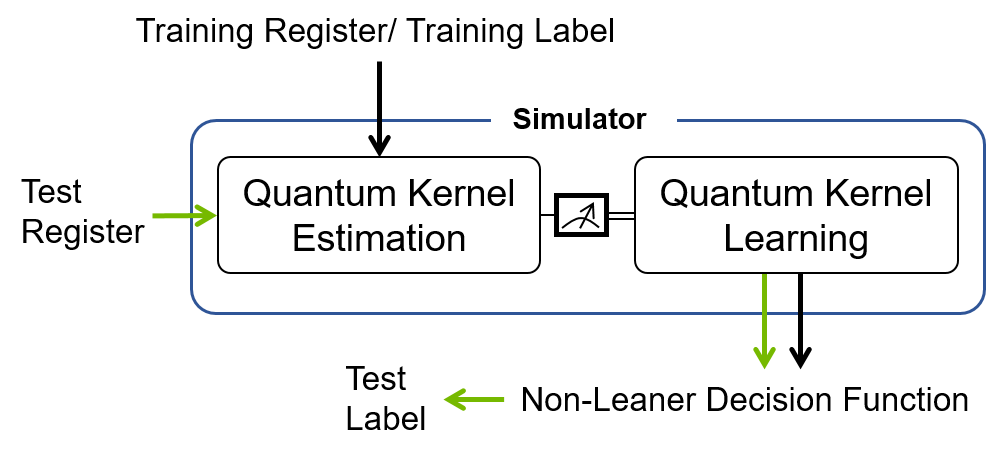}
    \caption{QSVM Simulator: Optimizes quantum kernel estimation and learning, enhancing phase operation and objective evaluation, leading to swift and precise classification outcomes.}
    \label{fig:qsvm_overview}
\end{figure}

Within the scope of our research, we have engineered an advanced tensor-network simulation framework, purpose-built to expedite the development of QSVMs through the integration of the cuTensorNet library underlying cuQuantum SDK\cite{bayraktar2023cuquantum}. This library is meticulously optimized for NVIDIA GPUs and can facilitate QSVM algorithms, requiring noiseless simulations for quantum kernel estimation as depicted in Fig. \ref{fig:qsvm_overview}. A pre-computation mechanism is embedded within this workflow, allowing for the reuse of an optimized tensor-network contraction path in the QSVM's complex learning stages, thereby bolstering the efficacy of both the training and classification phases.

Our tensor-network-based simulation is designed for parallel execution using the Message Passing Interface (MPI) and leverages the substantial computational power of GPU acceleration. This combination enables our QSVM simulator to efficiently manage large datasets while only modestly increasing memory requirements, thereby avoiding out-of-memory (OOM) situations during large-scale quantum circuit simulations.  Its flexibility ensures its utility across various quantum machine learning paradigms. Benchmark results show that our simulator achieves speedups often exceeding an order of magnitude compared with existing methods \cite{jones2019quest,gangapuram2024benchmarking}, thereby underscoring its potential as a robust and scalable tool for quantum machine learning within the broader Quantum-High-Performance Computing (HPC) ecosystem \cite{bayraktar2023cuquantum,chen2024quantum}.

A key feature of our simulator is its capacity to handle up to 784 qubits, enabling an extensive scaling analysis of QSVM performance and shedding light on the potential of quantum kernel methods in realistic data classification scenarios. Furthermore, this approach is flexible enough to accommodate various quantum machine learning paradigms and can be extended to multi-GPU settings for large-scale simulations. By validating our methods on real-world datasets (such as MNIST and Fashion-MNIST), we demonstrate that QSVMs can tackle complex classification tasks in Quantum-HPC environments, marking a significant step toward practical quantum-enhanced machine learning. These strides in QSVM development signal a major progression towards practical deployment, charting a path for the application of quantum-enhanced methodologies to complex, real-world data classification challenges within the Quantum-HPC Ecosystem\cite{chen2023quantum,chen2024quantum,vasques2023application,yang2023quantum,wu2021application}.

These results highlight not only the viability of QSVM algorithms but also the value of advanced simulation tools in guiding future quantum hardware development, such as offering ground truth for benchmarking purpose. As such, this work contributes to bridging the gap between theoretical QSVM formulations and their eventual implementation on large-scale quantum devices, offering valuable insights for both algorithmic refinement and hardware optimization in the quantum information sciences.

\section{Background}
QSVMs represent a significant breakthrough in quantum machine learning, particularly for large-scale data classification. The pioneering work by Rebentrost \emph{et al.}~\cite{rebentrost2014quantum} introduced a quantum algorithm that substantially enhances the computational efficiency of traditional SVMs. By harnessing quantum-mechanical principles such as superposition and interference, QSVMs can, under certain assumptions (e.g., quantum random-access memory, qRAM), achieve near-logarithmic complexity with respect to both the dimensionality of feature vectors $N$ (qubit number) and the size of the training dataset $M$ (data size). This approach suggests a potential exponential speedup over classical methods, although practical constraints like the realization of qRAM remain a major challenge.

More recent work on quantum machine learning has shifted toward quantum kernel estimation, emphasizing the capability of entangled quantum states to embed classical data in an exponentially large Hilbert space~\cite{havlivcek2019supervised}. Rather than focusing solely on matrix-inversion routines, these methods evaluate inner products of quantum states (i.e., kernel functions) that would be prohibitively expensive to compute classically. By embedding data into high-dimensional quantum feature spaces, one can construct decision boundaries that may be unreachable with purely classical algorithms. Indeed, Ref.~\cite{liu2021rigorous} demonstrates an end-to-end quantum speedup for a suitably constructed classification problem, providing concrete evidence that quantum kernels can yield practical gains in machine learning tasks.

Classical SVMs aim to find a hyperplane that maximizes the margin between two classes, typically formulated in its primal form as:

\begin{equation} \min_{\mathbf{w}, b} \frac{1}{2} ||\mathbf{w}||^2 \end{equation} subject to \begin{equation} y_\text{j}(\mathbf{w} \cdot \mathbf{x}_\text{j} + b) \geq 1, ; \forall \text{j}, \end{equation}

where $\mathbf{w}$ is the normal vector to the hyperplane, $b$ is the bias, $\mathbf{x}_j$ are feature vectors, and $y_j$ are the class labels. In quantum extensions of this method, the data are mapped into a higher-dimensional Hilbert space via a quantum kernel, enabling efficient non-linear classification that would otherwise be computationally prohibitive on classical hardware.

Early QSVM formulations relied on quantum matrix-inversion routines, such as the HHL algorithm~\cite{duan2020survey}, to mitigate the computational bottleneck inherent to large-scale SVMs. Theoretical analyses indicated that QSVM could perform these matrix inversions with $\mathcal{O}(\log(NM))$ complexity~\cite{rebentrost2014quantum}, a significant improvement over classical approaches. Quantum parallelism further reduces computational overhead by allowing simultaneous calculation of many kernel matrix elements, crucial for SVM optimization.

Within the QSVM framework, data points $x_\text{i}$ are non-linearly transformed into quantum states $\rho(x_\text{i}) = |\psi(x_\text{i})\rangle \langle \psi(x_\text{i})|$ within the Hilbert space. The inner product between these quantum states, crucial for constructing the kernel matrix $K(x_\text{i}, x_\text{j})$, is given by:

\begin{equation} K(x_\text{i}, x_\text{j}) = \text{tr}{\rho(x_\text{i})\rho(x_\text{j})} = |\langle \psi(x_\text{i}) | \psi(x_\text{j}) \rangle|^2, \end{equation}

where $|\langle \psi(x_\text{i}) | \psi(x_\text{j}) \rangle|^2$ is computed using a unitary matrix $U$, defined as:

\begin{equation} |\langle \psi(x_\text{i}) | \psi(x_\text{j}) \rangle|^2 = |\langle 0^{\otimes N} | U^\dagger(x_\text{i}) U(x_\text{j}) | 0^{\otimes N} \rangle|^2, \end{equation}

with $|0^{\otimes N} \rangle$ representing the initial state with all qubits in the $|0 \rangle$ state.

QSVM extends classical SVM by utilizing quantum superposition and entanglement to address large, complex datasets more efficiently. Quantum parallelism enables rapid evaluation of kernel functions across multiple data pairs, a task that is computationally expensive classically~\cite{gentinetta2024complexity}. In light of these developments, modern QSVM research increasingly emphasizes quantum kernel methods, reflecting both the capabilities of near-term quantum devices and the desire to circumvent the strict requirements of fully functional qRAM. Consequently, the present work examines classical simulations of quantum kernel approaches, building on recent theoretical and experimental advances to investigate whether and how quantum-enhanced feature spaces can yield advantages in realistic classification scenarios.

\section{Simulating QSVM with Tensor Networks Using the cuQuantum SDK and cuTensorNet Library}
\label{section.iii}

\subsection{Introduction of cuQuantum SDK and cuTensorNet Library}

\begin{figure}[!t]
    \centering
    \includegraphics[width=\textwidth]{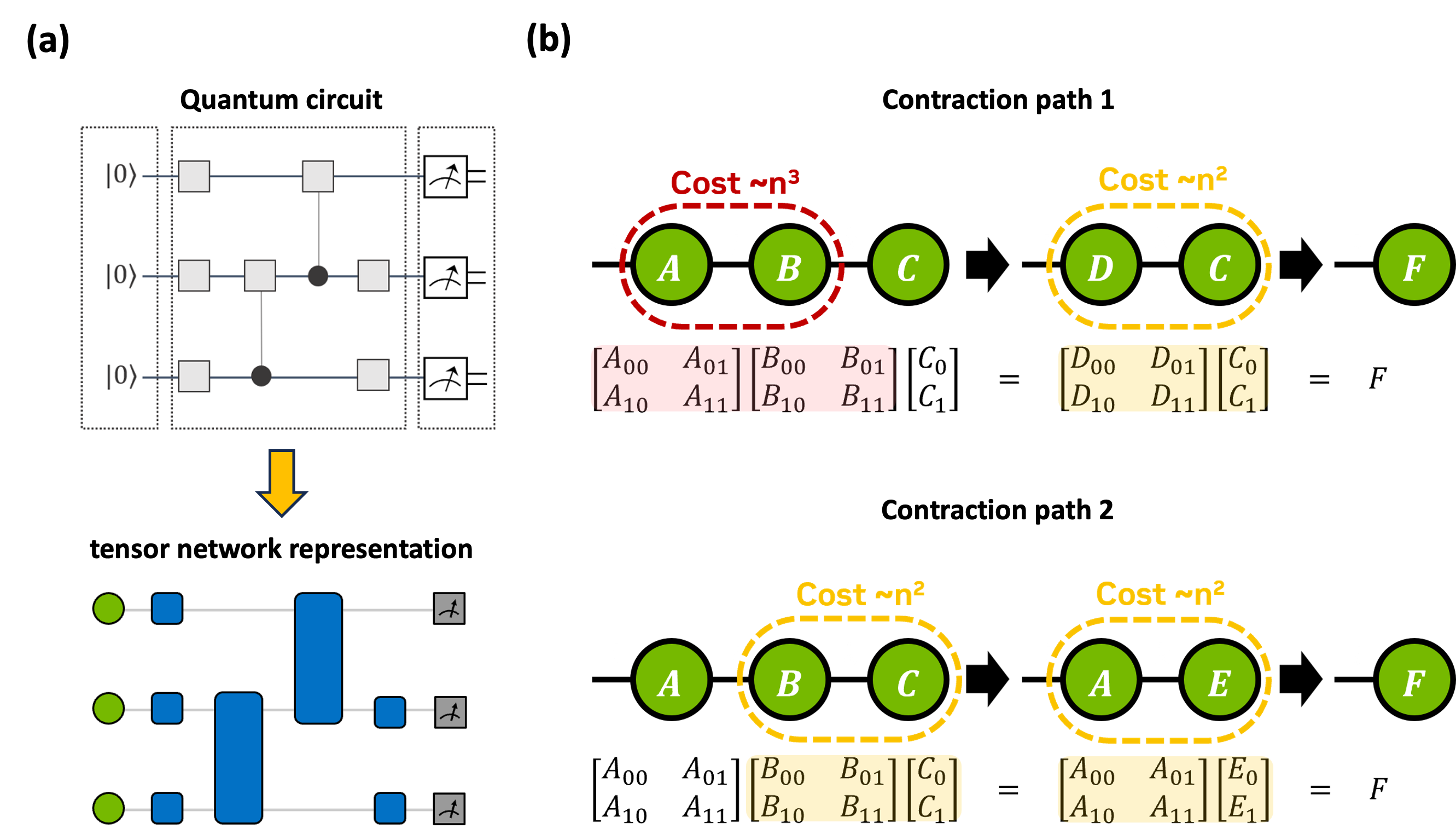}
    \caption{(a) Tensor network formulation of a quantum circuit. (b) Contraction paths determine the computational and memory costs of tensor network simulations: The upper path results in higher costs.}
    \label{fig:cutensornet_circuit}
\end{figure}

As the fields of quantum computing and advanced numerical simulations rapidly expand, NVIDIA has introduced cuQuantum SDK\cite{bayraktar2023cuquantum}, a comprehensive software development kit (SDK), to accelerate quantum circuit simulations with NVIDIA GPUs. It supports the programming model CUDA Quantum\cite{bayraktar2023cuquantum,kim2023cuda} (CUDA-Q), and frameworks like Qiskit\cite{wille2019ibm}, Pennylane\cite{bergholm2018pennylane}, and Cirq\cite{isakov2021simulations}. By offering a scalable and high-performance platform for quantum simulations, cuQuantum can democratize access to quantum computing research and even propel the field towards achieving real-world quantum machine learning applications. 

The cuQuantum SDK consists of optimized libraries such as cuStateVec and cuTensorNet. cuStateVec is dedicated to state-vector simulation methods, providing significant acceleration and efficient memory usage, while cuTensorNet focuses on tensor-network simulations. For tensor-network methods, the quantum circuit is initially converted into a tensor-network representation (Fig.\ref{fig:cutensornet_circuit}(a)). Subsequently, pairwise contraction paths are optimized to minimize computational complexity and memory footprint, followed by the execution of the computation. As shown in Fig.\ref{fig:cutensornet_circuit}(b), the sequence of pairwise contractions plays a role in computational cost. cuTensorNet efficiently identifies high-quality contraction paths\cite{bayraktar2023cuquantum}, accelerating quantum machine learning exploration, especially for high-dimensional data. The library offers advanced features like path optimization, approximate simulations, multi-GPU, and multi-node execution, enabling large-scale simulations and significantly advancing research into complex quantum algorithms across quantum physics, quantum chemistry, and quantum machine learning.

\begin{figure}[!b]
    \centering
    \includegraphics[width=0.85\textwidth]{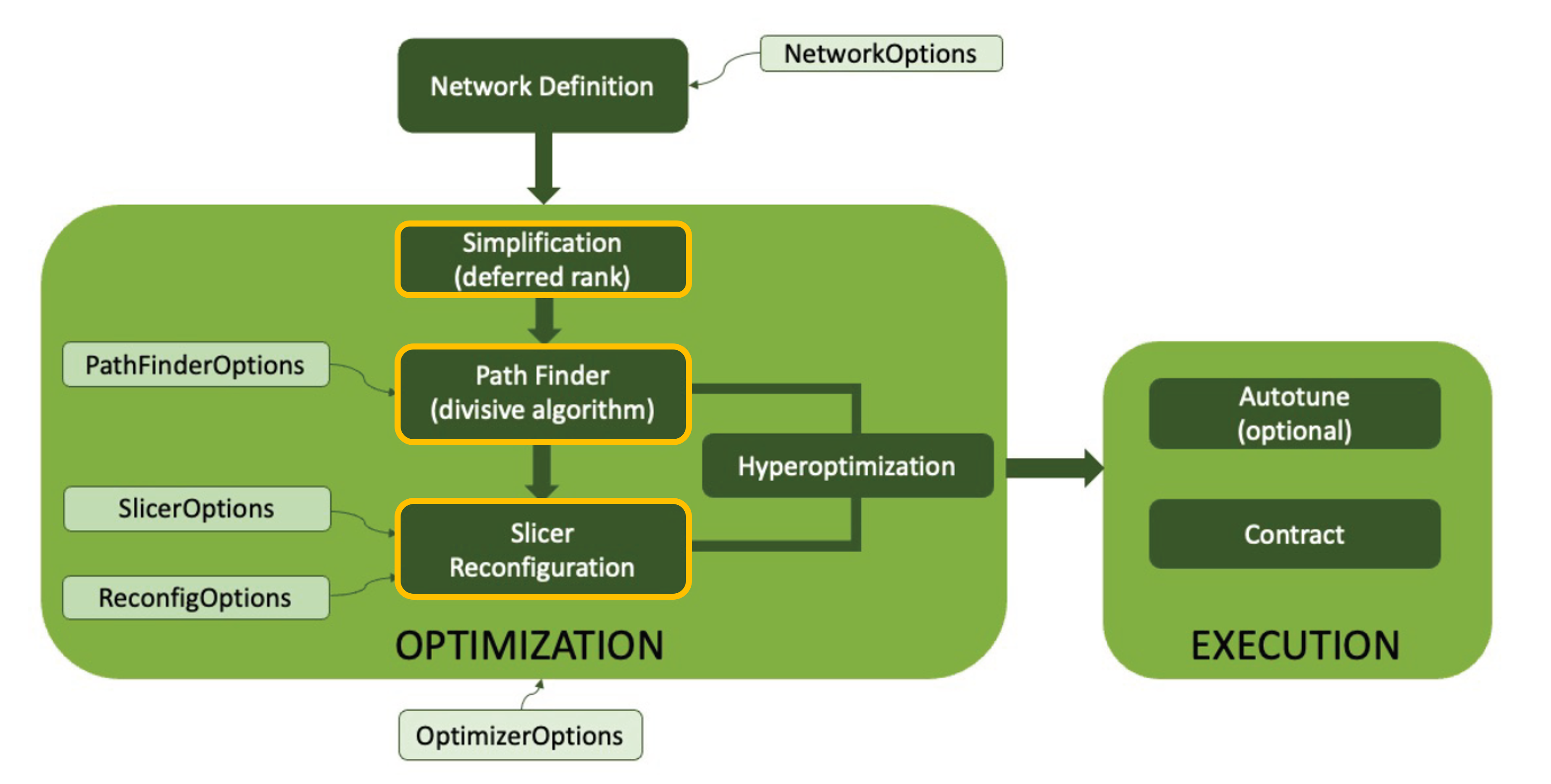}
    \caption{Building blocks of the cuTensorNet library.}
    \label{fig:cutensornet_module}
\end{figure}

To boost the efficiency of tensor network computation, cuTensorNet delivers modular and finely adjustable APIs, as shown in Fig. \ref{fig:cutensornet_module}, tailored for optimizing the pairwise contraction path on the CPU and improving contraction performance on the GPU. This optimization is essential for minimizing both computation cost and memory footprint. The pathfinder workflow is primarily structured in the following manner:

  \textit{1) Simplification:} This initial stage focuses on reducing the complexity of the whole tensor network and eliminating redundancies within the network. The implementation involves rank simplification to minimize the number of tensors by removing trivial tensor contractions from the network, resulting in a smaller network for subsequent processing.
  
  \textit{2) Division:} After simplification, the tensor network undergoes a recursive graph partitioning. This approach segments the network into multiple sub-networks and forms a contraction tree. The binary tree defines the contraction path and can be further optimized at the following stage.
  
  \textit{3) Slicing and Reconfiguration:} The slicing process selects a subset of edges from a tensor network for explicit summation. This technique results in lower memory requirements and allows parallel execution for each sliced contraction. Reconfiguration considers several small subtrees within the full contraction tree and attempts to reduce the contraction cost of the subtrees. cuTensorNet implements dynamic slicing, which interleaves slicing with reconfiguration.

\subsection{Pipeline of QSVM simulation}

\begin{figure}[!t]
    \centering
    \includegraphics[width=\textwidth]{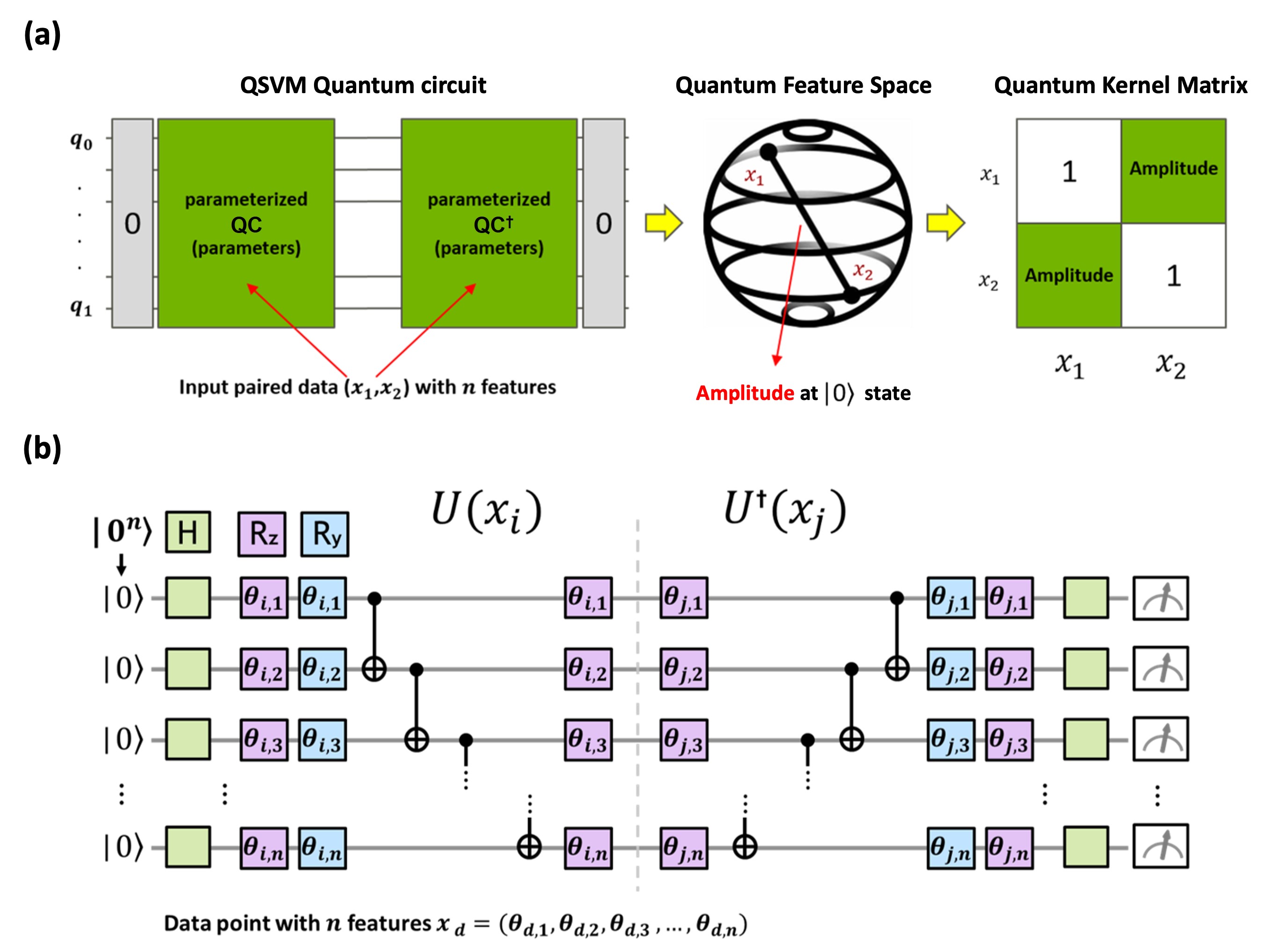}
    \caption{(a) The QSVM pipeline showcases the quantum circuit transformation of input data into feature space quantum states. (b) A schematic of the QSVM circuit.}
    \label{fig:pipeline}
\end{figure}

In Fig. \ref{fig:pipeline}(a), the depicted pipeline of a QSVM commences with the initial quantum state preparation in a canonical basis state \( |0\rangle \). The number of qubits depends on input data features, which can be adjusted using principal components analysis (PCA) to evaluate QSVM with varying qubit counts. The QSVM circuit comprises a parameterized quantum circuit (QC) and its corresponding adjoint (\(\text{QC}^\dagger\)), which correspond to the unitary operators \( U(\text{x}_i) \) and \( U^{\dagger}(\text{x}_j) \) depicted in Fig.~\ref{fig:pipeline}(b). The paired input data $\text{x}_i$ and $\text{x}_j$ are embedded into the left and right halves of the parameterized quantum circuit (QC and QC\(^\dagger\)), as shown in Fig.~\ref{fig:pipeline}(b). At the measurement stage, the probability amplitude of the zero state \( |0\rangle \) is used to represent the similarity between $\text{x}_i$ and $\text{x}_j$ in the quantum feature space. After computing the zero state amplitude for all paired data in the quantum feature space, the quantum kernel matrix is used to train the support vector classifier. Notably, only the probability of the all-zero state needs to be computed, allowing the tensor network simulation to reduce the overall computation by contracting the subspaces of the complete tensor structure. In this paper, we use a parameterized quantum circuit based on Block-Encoded State (BPS) wavefunctions\cite{martyn2020entanglement,suzuki2023quantum}. This enables QSVM to maintain high classification accuracy even with a greater number of qubits. Notably, the circuit does not decompose into smaller blocks; instead, each qubit is entangled through CNOT gates arranged in a linear topology, ensuring compatibility with near-term quantum hardware.

\subsection{Complexity of Quantum Circuit Simulation for QSVM}
\begin{figure}[htpb]
    \centering
    \includegraphics[width=\textwidth]{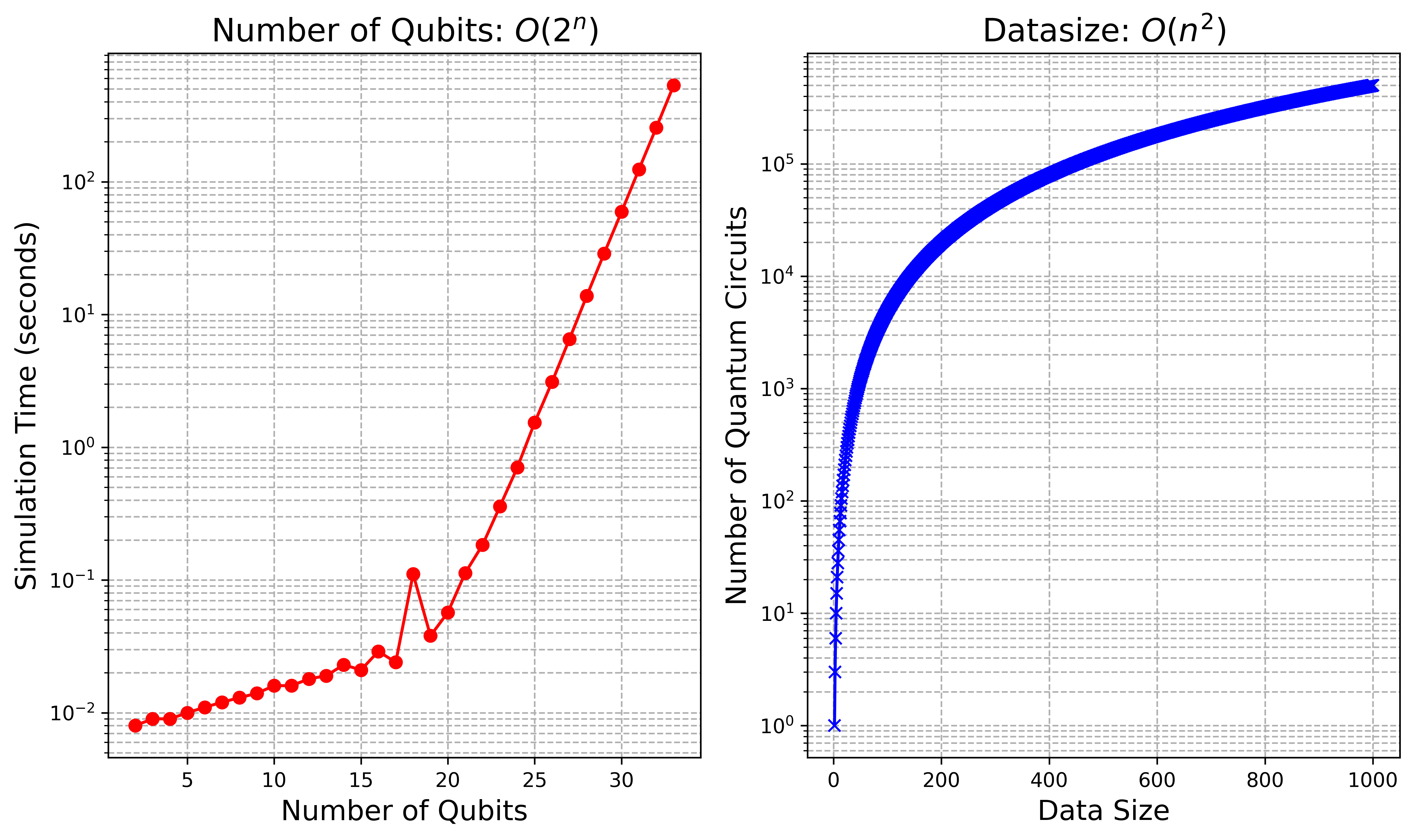}
    \caption{Computational complexity of QSVM simulation. The left graph demonstrates that simulation time scales exponentially with the number of qubits, as \(O(2^n)\), while the right graph shows that the number of quantum circuits required scales quadratically with data size, as \(O(n^2)\).}
    \label{fig:num_circuit}
\end{figure}

When executed on classical hardware such as CPUs and GPUs, the simulation of the QSVM algorithm poses significant computational challenges with state vector simulations. Figure \ref{fig:num_circuit} elucidates these challenges, indicating that the complexity scales exponentially with the number of qubits as \(O(2^n)\) and quadratically with data size as \(O(N^2)\). Additionally, the memory footprint of the full state vector grows exponentially with the number of qubits \(q \), which map features into Hilbert space for quantum circuit simulations. This aspect underscores the inherent computational intensity of simulating quantum systems on classical infrastructure. 

This scenario highlights the computational complexity advantages that QSVM offers in the realm of quantum machine learning. The simulation demands, in terms of computation time and memory size, grow exponentially with larger datasets and a greater number of qubits, a limitation not encountered when QSVM is run on quantum computers. As demonstrated by Rebentrost et al. \cite{rebentrost2014quantum}, the complexity advantage of QSVM can exhibit logarithmic scaling with respect to the product of the number of features and the size of the training set, denoted as \( O(\log(NM)) \). However, in the NISQ era, the verification of algorithms using traditional CPUs is inevitable. Therefore, this section focuses on leveraging GPU acceleration to address the computational bottlenecks encountered when simulating QSVM with large-scale qubit sizes and processing large datasets.

\subsection{Simulating QSVM's Quantum Kernel Matrix}
\label{section.iii(psudocode)}
In this section, we discuss three methods for simulating a QSVM algorithm: state-vector simulation on GPU using the cuStateVec library, tensor-network simulation on CPU using the opt-einsum library, and tensor-network simulation on GPU using the cuTensorNet library. In this research work, our interest lies in comparing a state-of-the-art CPU-centric approach (opt-einsum) to a state-of-the-art GPU-centric approach (cuTensorNet). This comparison highlights how GPU acceleration can significantly impact large-scale quantum circuit simulations.

\subsubsection{Simulation of QSVM with state vector}
\mbox{} \\
State vector simulation is widely used for simulating quantum circuit-based quantum computing because quantum circuit operations can naturally be represented using state vectors. In this method, Qiskit is used to create the QSVM quantum circuit, and the state-vector simulation is implemented on a GPU via the cuStateVec backend, as described in Algorithm \ref{algo:custatevec}. The advantage of using cuStateVec includes a speedup of the simulation time by leveraging GPU capabilities and enabling multi-GPU processing with MPI for distributed computing. The effectiveness of cuStateVec in enhancing quantum-circuit-simulation efficiency is evidenced in Lykov et al.'s research work using cuStateVec and the cuQuantum SDK\cite{lykov2023fast}.

\begin{algorithm}[!t]
\SetAlgoLined
\SetKwInOut{Input}{Input}
\SetKwInOut{Output}{Output}
\Input{Number of data1 $datasize1$, Number of data2 $datasize2$, List of quantum circuits $circuits$, Index of data1 and data2 combinations $indices$, statevector simulator $simulator$}
\caption{Get Kernel Matrix using cuStateVec}
\label{algo:custatevec}
\BlankLine
\begin{enumerate}
    \item Initialize $kernel\_matrix \in \mathbb{C}^{datasize1 \times datasize2}$ with all elements set to zero.
    \item Set the current operand index $i$ to $-1$.
    \item \For{$i_{1}, i_{2} \in \{1, \ldots, indices\}$}{
        \begin{enumerate}
            \item Update the circuits index $i \leftarrow i + 1$.
            \item Save circuits[i] statevector.
            \item Set transpile(circuits[i], simulator).
            \item Run simulator and save result $result$.
            \item Compute amplitude $amp \leftarrow \text{result.get\_statevector()}$.
            \item Calculate and store $kernel\_matrix[i_{1}-1][i_{2}-1] \leftarrow (\sqrt{\text{amp.real}^2 + \text{amp.imag}^2})$.
        \end{enumerate}
    }
    \item Symmetrize $kernel\_matrix$ by adding its transpose and an identity matrix: $kernel\_matrix \leftarrow kernel\_matrix + kernel\_matrix^T + \text{diag}(\mathbb{1}_{datasize1})$.
\end{enumerate}
\Return{$kernel\_matrix$}
\end{algorithm}

\subsubsection{Simulation of QSVM with tensor network}
\mbox{} \\
Even with \texttt{cuStateVec} enabling GPU acceleration, challenges persist due to the complexity of encoding the number of qubits \( O(2^n) \) and the size of the data \( O(n^2) \). To surmount these challenges, we present an innovative approach using the \texttt{cuTensorNet} library for QSVM simulation. In the creation of the tensor network representation, we seamlessly integrate Qiskit and cuQuantum's built-in \texttt{CircuitToEinsum} object. 

Initially, Qiskit is used to construct a \texttt{QuantumKernel} circuit, which is then transformed into ``expression" and ``operand" components by \texttt{CircuitToEinsum}. Due to the identical topological structure of the quantum circuit, the same ``expression" component can be reused for subsequent pairs of data. Meanwhile, the ``operand" is updated with parameters from the previously created operand. This approach rapidly transitions data pairs into tensor networks and preserves computational efficiency. The derivation of the kernel matrix—a pivotal component of the SVM—exploits a consistent ``path" to greatly minimize the repetition of contraction order calculations. The detailed algorithm is described in Algorithm \ref{algo:cutensornet}. This technique not only leverages the computational strength of GPUs but also ensures path reusability, resulting in a considerable acceleration of the simulation process and a dramatic reduction in computational complexity. We will demonstrate those improvements in the next section. 

\begin{algorithm}[htpb]
\SetAlgoLined
\SetKwInOut{Input}{Input}
\SetKwInOut{Output}{Output}
\Input{Number of data1 $datasize1$, Number of data2 $datasize2$, Circuit einstein summation expression $exp$, List of operands $operands$, Index of data1 and data2 combinations $indices$, network options $options$}
\caption{Get Kernel Matrix using cuTensorNet} 
\label{algo:cutensornet}
\BlankLine
\begin{enumerate}
    \item Initialize $kernel\_matrix \in \mathbb{C}^{datasize1 \times datasize2}$ with all elements set to zero.
    \item Set the current operand index $i$ to $-1$.
    \item Initialize the network with given $options$ to prepare for contraction operations.
    \item \For{$i_1, i_2 \in \{1, \ldots, indices\}$}{
        \begin{enumerate}
            \item Update the operand index $i \leftarrow i + 1$.
            \item Reset the network to its initial state before each contraction.
            \item Prepare the operands for contraction based on $i$.
            \item Compute amplitude $amp \leftarrow \text{Contract within the network}(exp, operands[i], options)$.
            \item Calculate and store\\ $kernel\_matrix[i_1-1][i_2-1] \leftarrow \sqrt{\text{amp.real}^2 + \text{amp.imag}^2}$.
        \end{enumerate}
    }
    \item Symmetrize $kernel\_matrix$ by adding its transpose and an identity matrix:\\ $kernel\_matrix \leftarrow kernel\_matrix + kernel\_matrix^T + \text{diag}(\mathbb{1}_{datasize1})$.
\end{enumerate}
\Return{$kernel\_matrix$}
\end{algorithm}

To ensure a fair comparison tensor network QSVM simulation between CPU and GPU performance, we utilize the opt-einsum package, which provides optimized tensor computation on CPUs similar to the cuQuantum SDK available for NVIDIA GPUs. The detailed algorithm for simulating the QSVM on CPUs, aimed at equalizing the computational environment to the extent possible, is described in Algorithm \ref{algo:einsum}.

\begin{algorithm}[!t]
\SetAlgoLined
\SetKwInOut{Input}{Input}
\SetKwInOut{Output}{Output}
\Input{Number of data1 $datasize1$, Number of data2 $datasize2$, Circuit einstein summation expression $exp$, List of operands $operands$, Index of data1 and data2 combinations $indices$, Contraction path $path$}
\caption{Get Kernel Matrix using opt-einsum}
\label{algo:einsum}
\BlankLine
\begin{enumerate}
    \item Initialize $kernel\_matrix \in \mathbb{C}^{datasize1 \times datasize2}$ with all elements set to zero.
    \item Set the current operand index $i$ to $-1$.
    \item \For{$i_1, i_2 \in \{1, \ldots, indices\}$}{
        \begin{enumerate}
            \item Update the operands index $i \leftarrow i + 1$.
            \item Compute amplitude\\ $amp \leftarrow \text{opt\_einsum.contract}(exp, operands[i], path)$.
            \item Calculate and store\\ $kernel\_matrix[i_1-1][i_2-1] \leftarrow \sqrt{\text{amp.real}^2 + \text{amp.imag}^2}$.
        \end{enumerate}
    }
    \item Symmetrize $kernel\_matrix$ by adding its transpose and an identity matrix: $kernel\_matrix \leftarrow kernel\_matrix + kernel\_matrix^T + \text{diag}(\mathbb{1}_{datasize1})$.
\end{enumerate}
\Return{$kernel\_matrix$}
\end{algorithm}

\section{Performance and Benchmarking of QSVM with cuTensorNet}
\subsection{QSVM Simulation and cuTensorNet-Accelerated QSVM (cuTN-QSVM)}
In the outlined simulation workflow, Fig. \ref{fig:qsvm_overview} and \ref{fig:pipeline} illustrate the sequence from the initial input of data to the generation of a quantum circuit for the purpose of encoding. Subsequent steps involve the use of optimized compilation to compute and simulate the quantum circuits, leading to the extraction of a quantum kernel matrix. This matrix is then applied to develop a support vector classifier (SVC). 

\begin{figure*}[!t]
    \centering
    \includegraphics[width=0.9\textwidth]{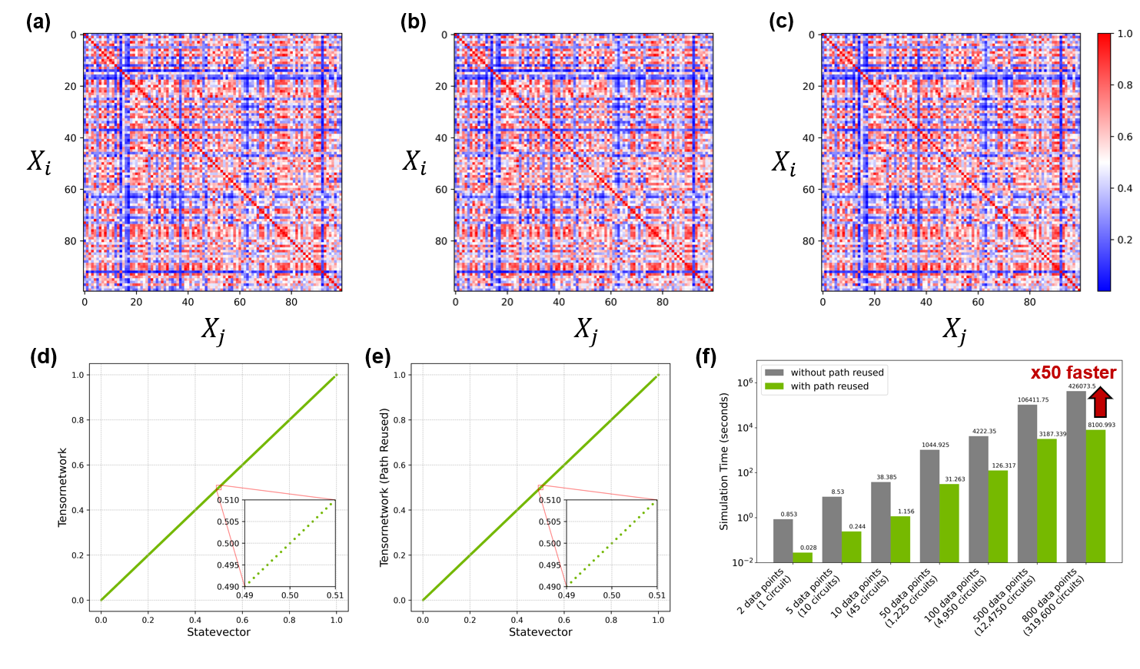}
    \caption{Comparative visualization of quantum kernel matrices and their computation speedups. (a), (b), and (c) illustrate the quantum kernel matrices generated from state vector simulation, tensor network simulation, and tensor network simulation with path reuse strategies, respectively. (d) and (e) feature the parity plots for quantum kernel assessments comparing the outputs of state vector simulations with tensor network and tensor network with path reuse algorithms, demonstrating high concordance. (f) quantifies the performance enhancement attributable to path reuse in tensor network simulations, showcasing significant temporal reductions across an array of dataset sizes.}
    \label{fig:kernelmatrix_compare}
\end{figure*}

However, in typical CPU-based workflows, bottlenecks arise in the progression from the construction of quantum circuits to the calculation of the quantum kernel matrix, where the complexity of simulating the QSVM algorithm scales exponentially with the number of qubits, \(O(2^n)\), and quadratically with data size, \(O(N^2)\). To alleviate these bottlenecks, we incorporate the cuQuantum SDK into the QSVM workflow, employing a method of assigned parameters for the formulation of QSVM's quantum circuits. We then maintain a consistent ``expression" for the simulation of these circuits. Ultimately, a ``path reuse" strategy is adopted for the tensor network contractions to compute the quantum kernel matrix, effectively reducing redundant computations when processing large datasets, reducing it from \(O(N^2)\) to \(O(1)\) for pathfinding. Importantly, as depicted in Fig. \ref{fig:kernelmatrix_compare}, the expressions and paths used in the cuTensorNet during the QSVM simulation process remain unchanged compared to those in CPU and cuStateVec, ensuring that no accuracy is compromised for the sake of expedience. In addition to the path reuse strategy, cuTensorNet offers concurrent execution for tensor network contractions. This technique allows the continued contractions on multiple GPUs asynchronously when tensors are already on the device, thus enhancing computational efficiency by continuing operations without delay.  The pronounced speedup achieved through the implementation of path reuse within the cuTensorNet library is detailed in Fig.\ref{fig:kernelmatrix_compare}(g), where we report a fiftyfold increase in speed compared to conditions without path reuse.

\begin{figure}[!t]
    \centering
    \includegraphics[width=0.75\textwidth]{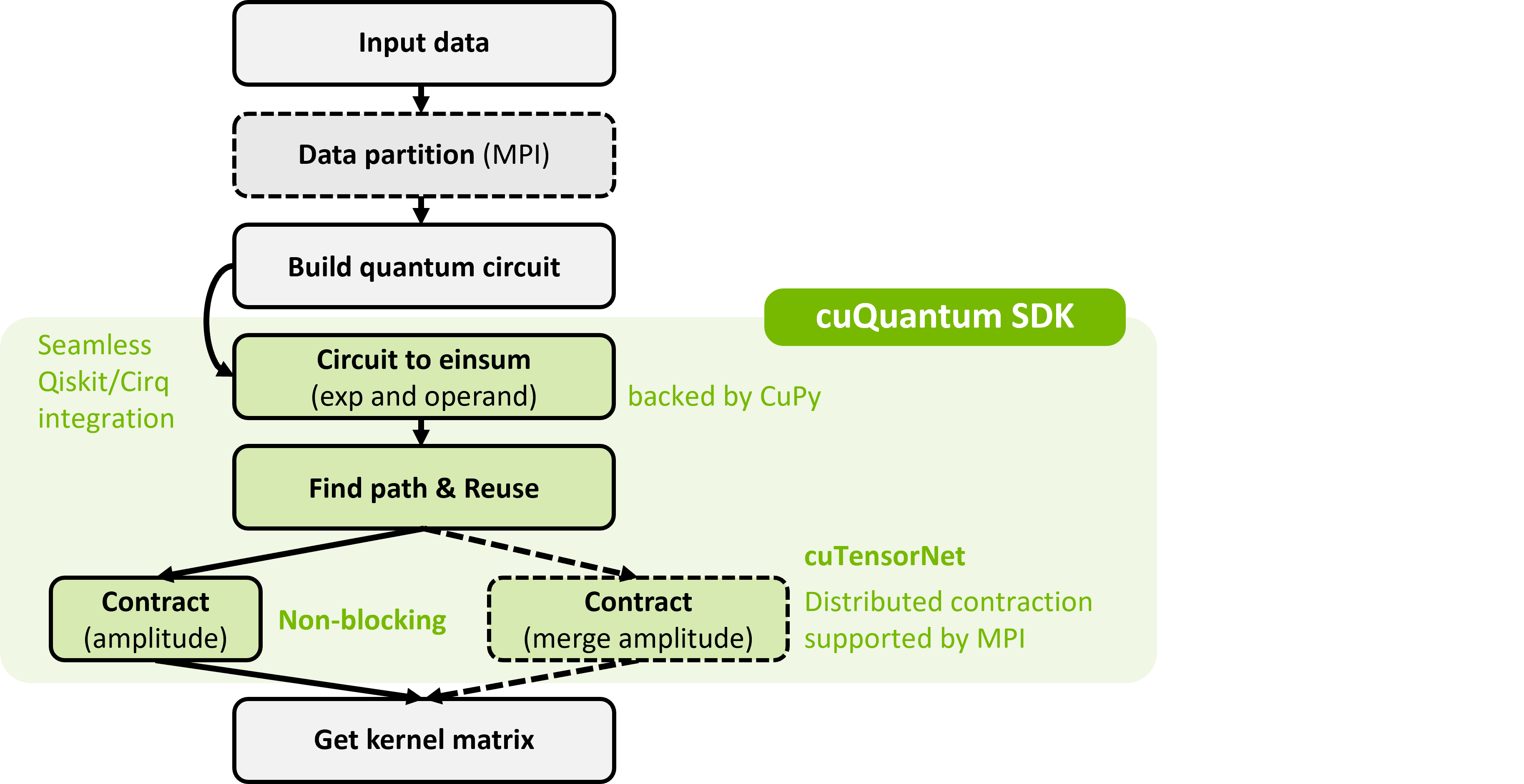}
    \caption{Workflow optimization for QSVM simulation through architectural enhancements, integrating Qiskit/Cirq with cuQuantum SDK. This transition from circuit building to tensor network conversion and kernel matrix computation reduces computational time complexity, leveraging GPU acceleration and multi-GPU strategies.}
    \label{fig:process_flow_comparison}
\end{figure}

In the comprehensive workflow outlined in Fig.~\ref{fig:process_flow_comparison}, the input data initiates quantum circuit construction, integrating frontends such as Qiskit or Cirq with the cuQuantum API, which generates Einstein summation expressions and tensor operands for the circuit. The process advances by converting quantum circuits into tensor networks represented as CuPy arrays, enabling the utilization of in-place operations to update content for the same operands efficiently. Key to enhancing computational efficiency within this framework is the strategic deployment of direct conversion from data to operand, alongside expression reuse for optimizing computational pathways. This step is crucial in minimizing redundancy and ensuring the streamlined execution of the workflow. As the process proceeds, CuPy's capabilities are harnessed to accelerate the computation of the kernel matrix, culminating in the application of the SVC. Moreover, cuTensorNet, as part of the cuQuantum SDK, incorporates advanced strategies such as path reuse and non-blocking operations across multi-GPU configurations. 

\begin{figure}[!b]
    \centering
    \includegraphics[width=0.65\textwidth]{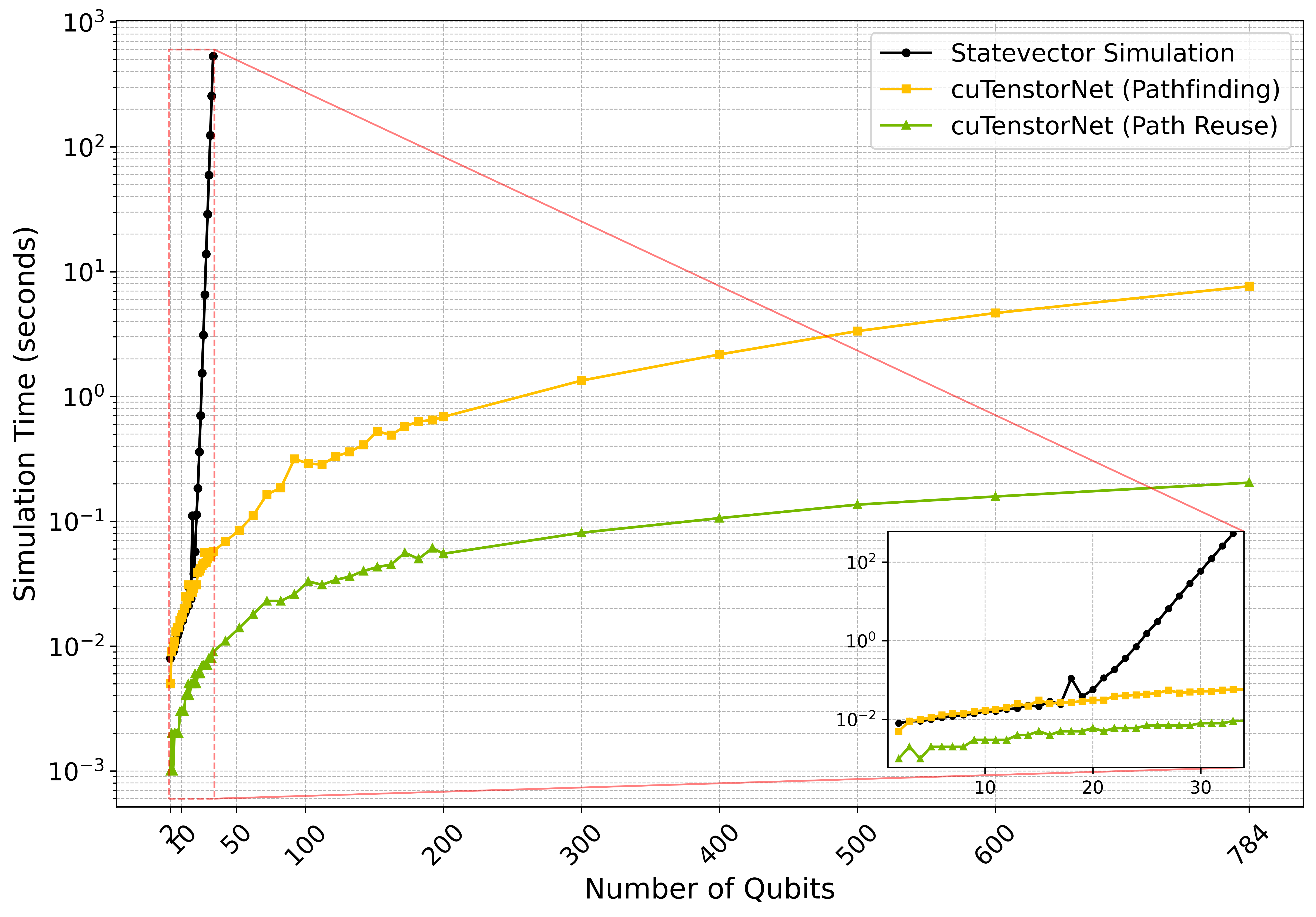}
    \caption{Simulation time comparison for quantum state vector simulation and two cuTensorNet approaches—path plus contraction (no-path reuse), and contraction only (path reuse)—across a range of qubit numbers, up to the equivalent of a 28x28 pixel grid (784 qubits). }
    \label{fig:acceleration}
\end{figure}

These approaches significantly reduce the computational overhead from a conventional complexity of $O(2^n)$ to a more scalable $O(n^2)$, thereby enhancing the practicality of executing extensive QSVM simulations with improved processing times and efficiency in resource utilization. Fig. \ref{fig:acceleration} illustrates that quantum simulation on the NVIDIA A100 GPU using cuStateVec becomes practically infeasible for more than 50 qubits. However, by employing cuTensorNet, single-contraction simulations can be completed within 0.2 seconds, even with up to 784 qubits. Additionally, Fig. \ref{fig:acceleration} shows that the path reuse strategy can further enhance the speed, offering more than tenfold acceleration when increasing the number of qubits in the QSVM algorithm.

In the GPU-accelerated workflow utilizing cuTensorNet, as delineated in Fig. \ref{fig:process_flow_comparison}, we are able to expand the feature size (number of qubits) and scale up the data volume for our QSVM algorithm. The evaluation of accuracy resulting from these augmentations will be discussed in the following part, while an in-depth assessment of resource management will be presented in the subsequent section.

\subsection{Accuracy Benchmarking and Validation of Large-scale QSVM}
\subsubsection{Binary Classification}
\mbox{} \\
We benchmarked the performance of a QSVM on high-dimensional image classification tasks using both 10-class MNIST and Fashion-MNIST datasets of 31,500 images each, split 80–20 for training and testing. As a classical baseline, we employed an SVM with a radial basis function (RBF) kernel and systematically varied the scaling parameter $\gamma$ between 0.001 and 1000 to identify optimal hyperparameters. Although SVM is not the most advanced machine learning model available, it provides a well-established framework that enables us to validate the feasibility and potential advantages of our quantum-enhanced approach.

\begin{figure}[!b]
    \centering
    \includegraphics[width=\textwidth]{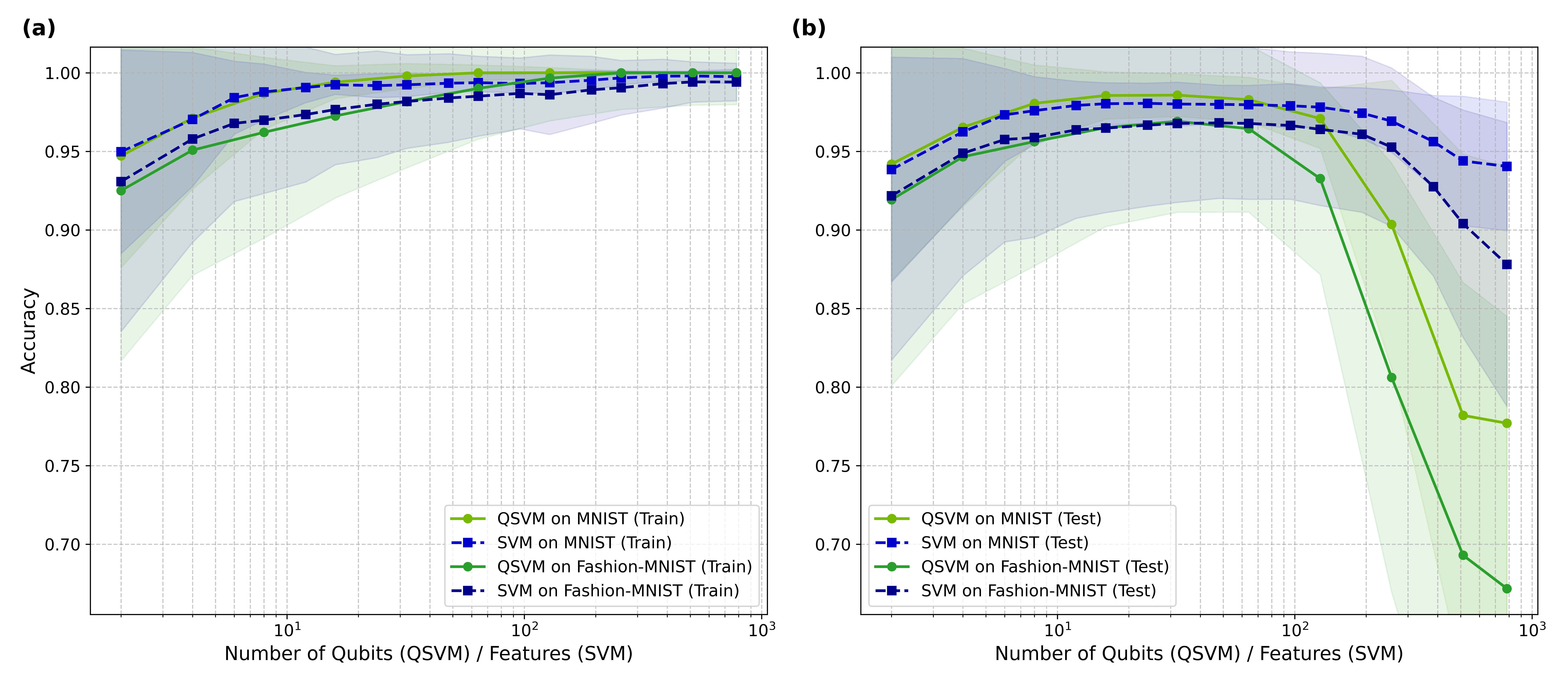}
    \caption{Benchmark of QSVM and SVM on MNIST and Fashion-MNIST (nine labels, 45 binary classification tasks). (a) Training accuracy vs number of qubits (QSVM) or features (SVM). (b) Test accuracy. QSVM outperforms SVM for moderate qubit counts, but overfitting leads to reduced test accuracy with more qubits. Shaded regions show one standard deviation around mean accuracy.}
    \label{fig:mnist-binary-result}
\end{figure}

In Fig.~\ref{fig:mnist-binary-result}, our results indicate that QSVM offers competitive performance and can, under certain conditions, outperform the classical SVM for moderate circuit sizes. In particular, QSVM maintains high accuracy by leveraging quantum feature maps that embed data into larger Hilbert spaces. However, beyond a critical qubit threshold, we observe a decline in test accuracy, which we attribute to overfitting effects and the phenomenon of Barren plateaus—where off-diagonal kernel matrix elements diminish and the optimization landscape becomes exponentially flat\cite{mcclean2018barren}. This vanishing-gradient problem not only complicates the training of parameterized quantum circuits but also underscores the practical limitations of naively scaling up circuit depth or qubit count.

\begin{figure}[htpb]
    \centering
    \includegraphics[width=0.75\textwidth]{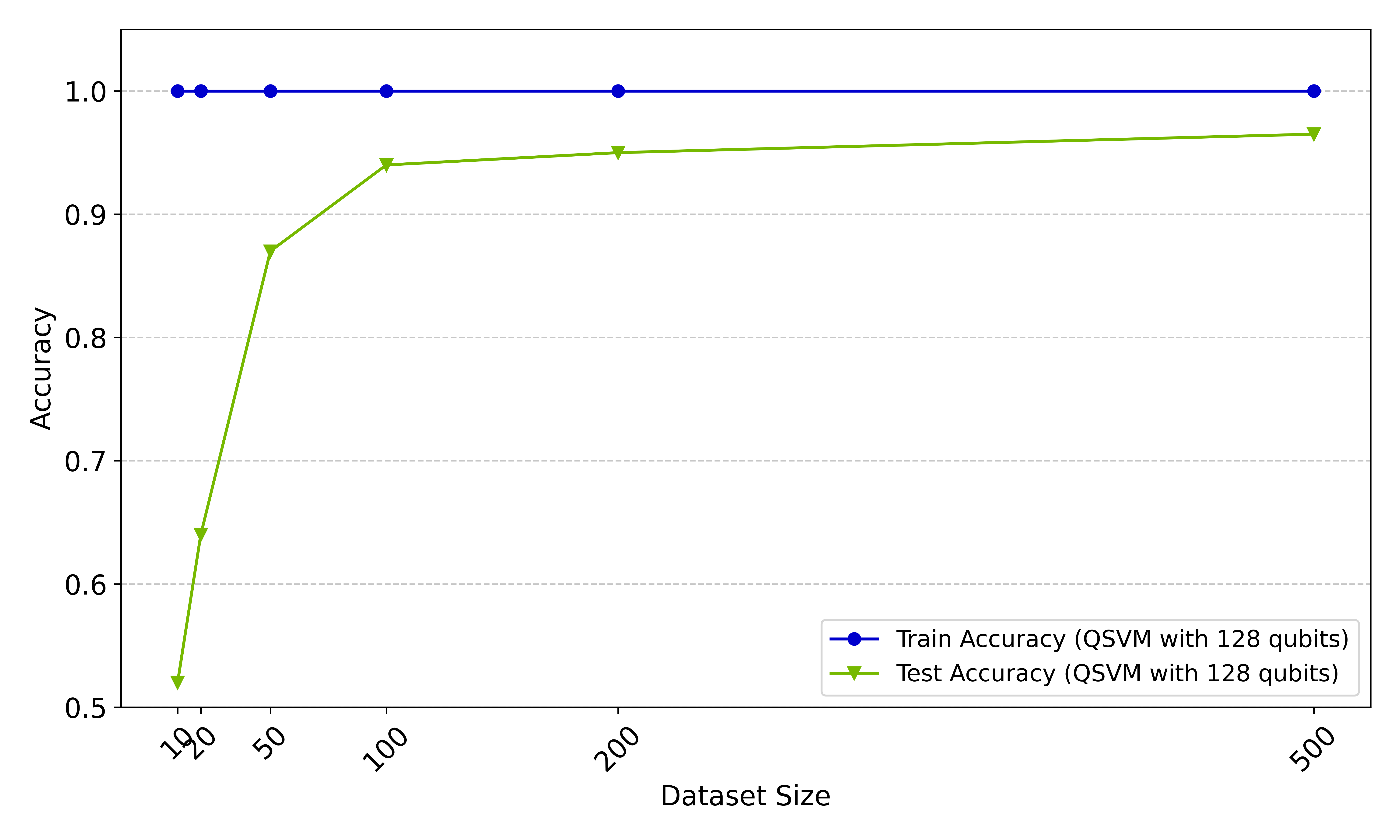}
    \caption{Classification Accuracy versus Dataset Size for Binary Classification of MNIST Digits 2 and 6 using a 128-Qubit QSVM model.}
    \label{fig:accuracy_data}
\end{figure}

In light of these observations, current quantum machine learning approaches still require careful feature engineering or hybrid methods to optimize model performance. Moreover, the amount of training data can significantly impact QSVM accuracy, as illustrated in Fig.~\ref{fig:accuracy_data}, underscoring the importance of sufficiently large datasets.

Despite these challenges, our proof-of-concept study confirms that QSVM can serve as a promising foundation for large-scale quantum machine learning, particularly in scenarios where high-dimensional embeddings may confer a computational advantage. Further research on quantum circuit design, regularization strategies, and optimization techniques will be crucial to fully harness the benefits of quantum-enhanced models and to mitigate the pitfalls associated with increasingly large quantum systems.

\subsubsection{Multi-class Classification}
\mbox{} \\
\begin{figure}[!b]
    \centering
    \includegraphics[width=\textwidth]{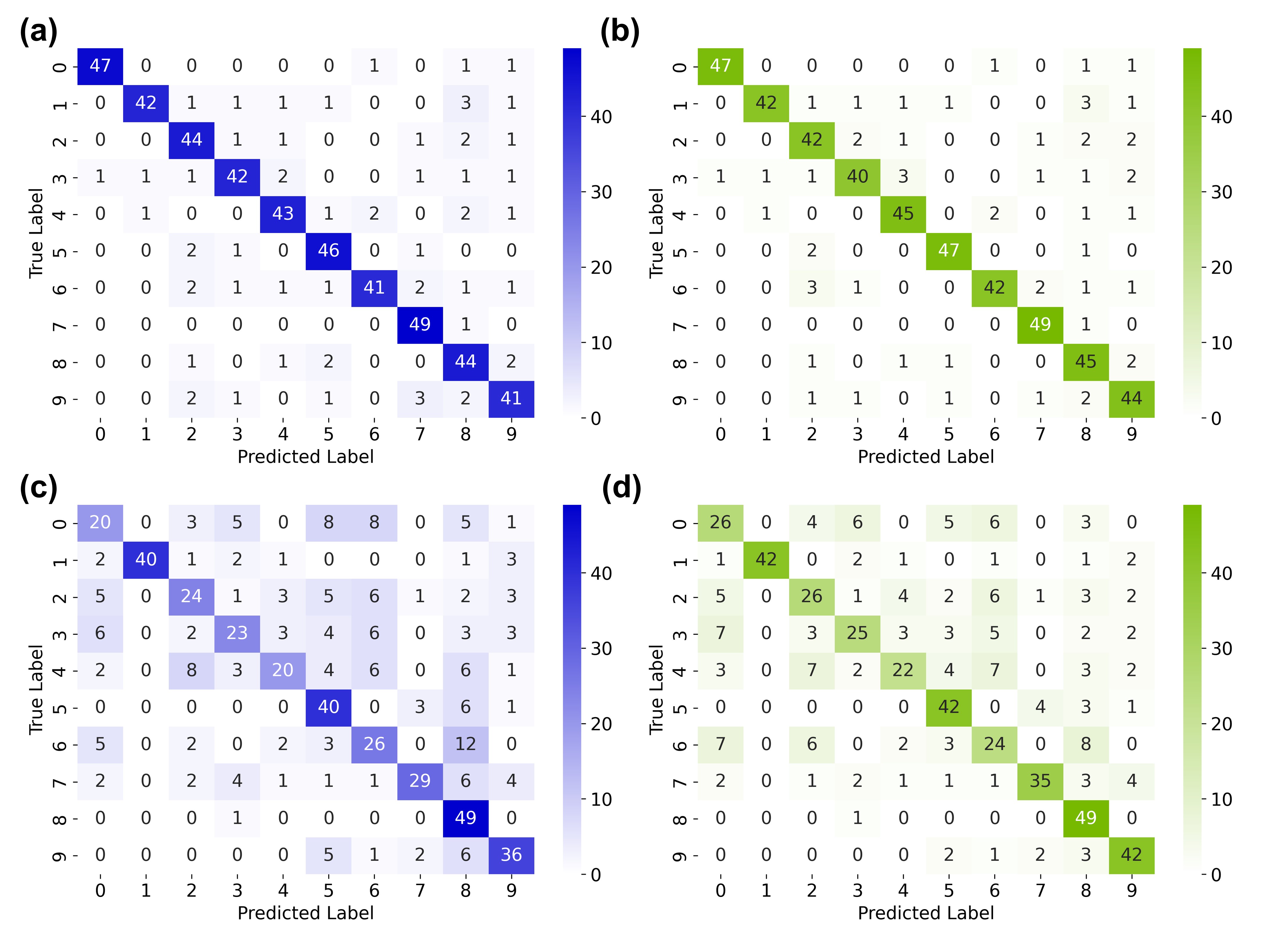}
    \caption{Confusion matrices for the classical SVM with 64 features (a) and 96 features (c), and for the QSVM with 64 qubits (b) and 96 qubits (d), evaluated on 10-class MNIST (a, b) and 10-class Fashion-MNIST (c, d). Each cell indicates the number of predictions for a given true label (vertical axis) and predicted label (horizontal axis). Brighter diagonal entries reflect a higher count of correctly classified samples. Overall, QSVM demonstrates consistently strong performance, often improving upon the classical SVM.}
    \label{fig:confusion_matrix}
\end{figure}

To assess the robustness of the QSVM beyond binary classification, we further evaluated its performance on the 10-class versions of the MNIST and Fashion-MNIST datasets. For each dataset, we select 1,000 training samples and reserve an additional 500 samples for testing. For MNIST, both the classical baseline SVM and the QSVM employ 64 features/qubits, whereas for Fashion MNIST, they use 96 features/qubits. Fig.~\ref{fig:confusion_matrix} shows the confusion matrices for both datasets under SVM and QSVM, while Table~\ref{tab:macro_metrics} summarizes several macro-level performance metrics (accuracy, sensitivity, specificity, and $F_1$ score). For MNIST, the QSVM yields slightly higher accuracy, along with marginal improvements in sensitivity and specificity. A similar trend emerges for the more challenging Fashion-MNIST dataset, where QSVM also achieves higher overall accuracy and macro-level metrics than the classical SVM.

These findings reinforce our earlier observations that, for moderate circuit sizes, QSVM can learn complex data distributions effectively. By embedding data points in a larger Hilbert space, the quantum kernel method can capture subtle features that improve class separability. However, as discussed previously, overfitting and the onset of Barren plateaus can degrade performance when the number of qubits becomes excessively large. Consequently, the design of circuit architectures and the choice of hyperparameters—particularly in multiclass settings—remain critical in balancing expressivity and generalization.

\begin{table}[ht]
\centering
\caption{Macro Metrics for SVM and QSVM on MNIST and Fashion MNIST Datasets}
\label{tab:macro_metrics}
\begin{tabular}{llcccc}
\toprule
\textbf{Dataset} & \textbf{Model} & \textbf{Accuracy} & \textbf{Sensitivity} & \textbf{Specificity} & \textbf{F1-score} \\
\midrule
\multirow{2}{*}{MNIST} & SVM   & 0.8780 & 0.8820 & 0.9865 & 0.8783 \\
                       & QSVM  & \textbf{0.8860} & \textbf{0.8900} & \textbf{0.9874} & \textbf{0.8864} \\
\midrule
\multirow{2}{*}{Fashion MNIST} & SVM   & 0.6140 & 0.6388 & 0.9578 & 0.6088 \\
                              & QSVM  & \textbf{0.6660} & 0\textbf{.6744} & \textbf{0.9633} & \textbf{0.6608} \\
\bottomrule
\end{tabular}
\end{table}

\subsection{Simulation with Single CPU and GPU}

\begin{figure}[!b]
    \centering
    \includegraphics[width=\textwidth]{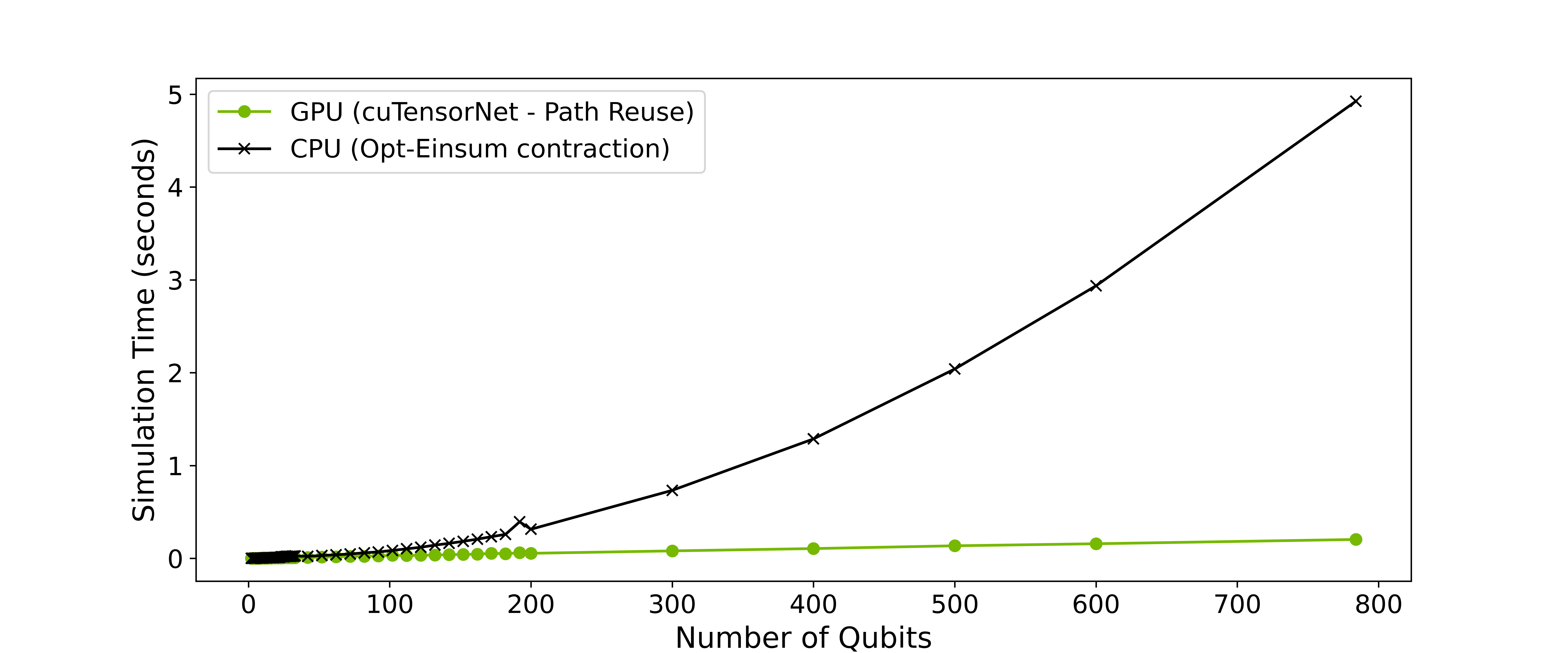}
    \caption{Benchmarking QSVM circuit simulation time using a single CPU and a single GPU. The GPU data shows minimal variation compared to the CPU's scale.}
    \label{fig: cpu_gpu_comparison}
\end{figure}

In this section, we compare the performance of a CPU and a GPU, as illustrated in Fig.\ref{fig: cpu_gpu_comparison}. To ensure a fair comparison, we employed Opt-Einsum for the contraction process on a single AMD EPYC 7J13 CPU, contrasting this with a single NVIDIA A100 GPU using cuTensorNet for the contraction process, with path reuse implemented. The detailed pseudocodes are discussed in Section \ref{section.iii(psudocode)}. Moreover, it was necessary to synchronize the contraction paths in Opt-Einsum with those of cuTensorNet to ensure consistency. As depicted in Fig.\ref{fig: cpu_gpu_comparison}, the speedup provided by the GPU relative to the CPU becomes more pronounced as the number of simulated qubits increases. Consequently, for large-scale qubit simulations, GPUs demonstrate enhanced scalability and promise substantial benefits for future advanced qubit algorithms in simulation and emulation.

\section{Distributed simulation and Resource Estimation in HPC}
\begin{figure}[!b]
    \centering
    \includegraphics[width=\textwidth]{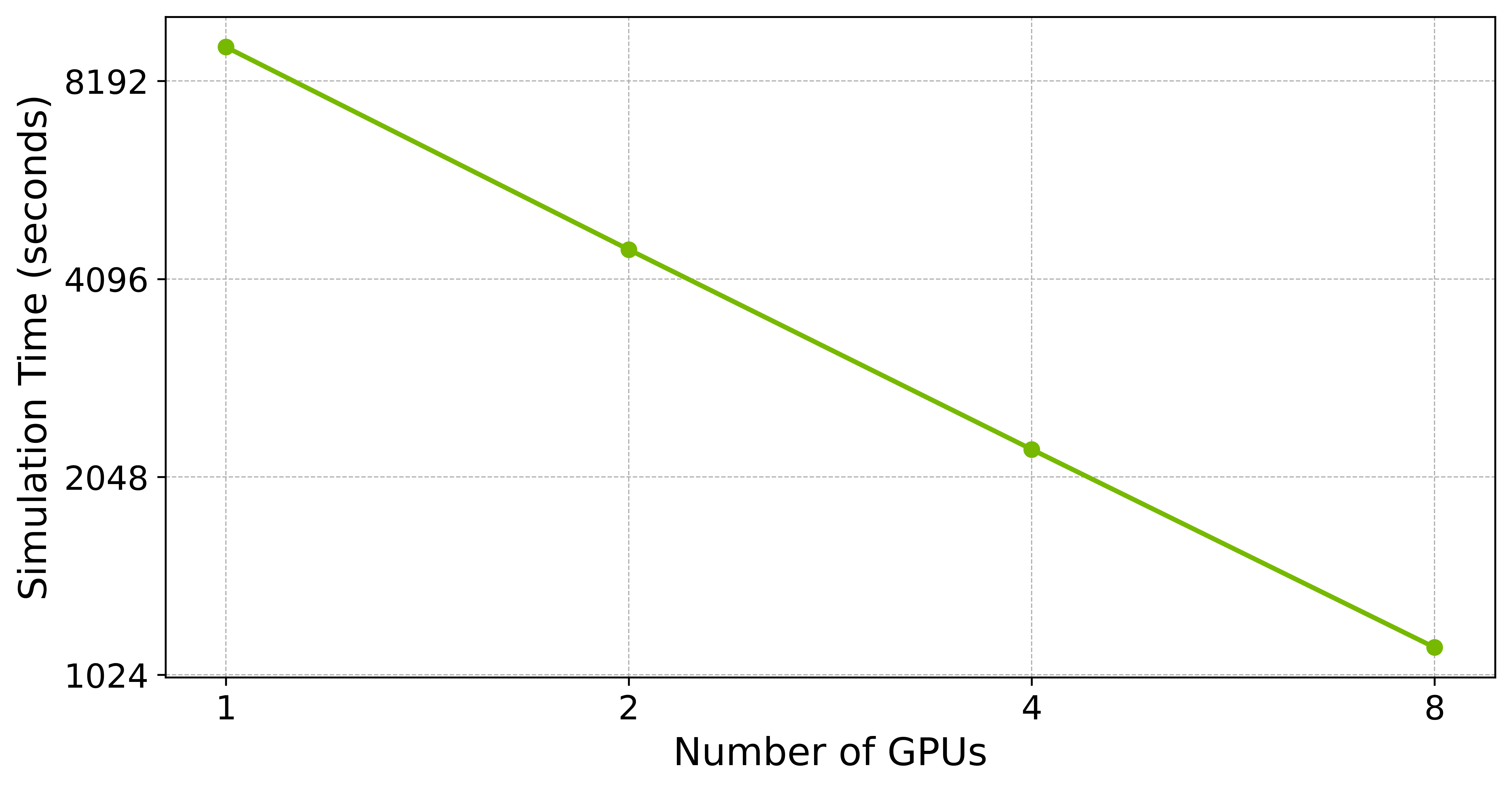}
    \caption{Strong scaling of the QSVM simulation is observed for 1,000 data points across 1, 2, 4, and 8 GPUs, demonstrating linear speedup.}
    \label{fig:multigpu}
\end{figure}

In the final section of our study, a multi-GPU instance was utilized to expand the QSVM model via cuTensorNet to accommodate a dataset comprising 1,000 data points of 28x28 MNIST images. The implementation of multi-GPU resources to enhance quantum circuit simulation via cuStateVec is thoroughly detailed in the research conducted by Shaydulin et al. \cite{shaydulin2023evidence}. Our emphasis lies on leveraging the data from these experiments to rigorously assess both the computational costs and the temporal demands inherent in the tensor-network simulation of the QSVM algorithm within a multi-GPU processing framework. 

In our computational environment, each GPU within a node is interconnected using the high-bandwidth NVLink network and the NVIDIA Collective Communications Library (NCCL) to optimize intra-node communication. The input data is paired and evenly distributed across multiple GPUs via NCCL, where it is directly converted into a tensor network for computation. The results are then returned to a single GPU via NCCL to form the quantum kernel matrix for SVC classification. By harnessing these integrated technological benefits, we have successfully actualized the accelerated computational outcomes for managing large-scale qubit systems and complex datasets, as illustrated in Fig.\ref{fig:multigpu}. Comparative analysis indicates that our performance metrics are on par with distributed simulation results documented in the existing scientific corpus, as cited in Bayraktar et. al.'s and Lykov et. al.'s work\cite{bayraktar2023cuquantum,lykov2023fast}.

\subsection{Benchmarking cuTensorNet Multi-GPU with MPI}
Fig. \ref{fig:multigpu} illustrates the execution time required for quantum simulations in relation to the number of qubits. The data compares the performance of a single A100 GPU to systems utilizing 2, 4, and 8 GPUs in conjunction with MPI and within a single NVIDIA DGX node. It is evident from the results that the incorporation of multi-GPUs significantly decreases computation time, highlighting the strong linear speedup of cuTenserNet with MPI. The trend indicates a substantial reduction in execution time as the number of GPUs is increased, affirming the efficacy of multi-GPU setups in handling large datasets.

\subsection{Large Dataset Processing with Multi-GPU}
\begin{figure}[!t]
    \centering
    \includegraphics[width=0.9\textwidth]{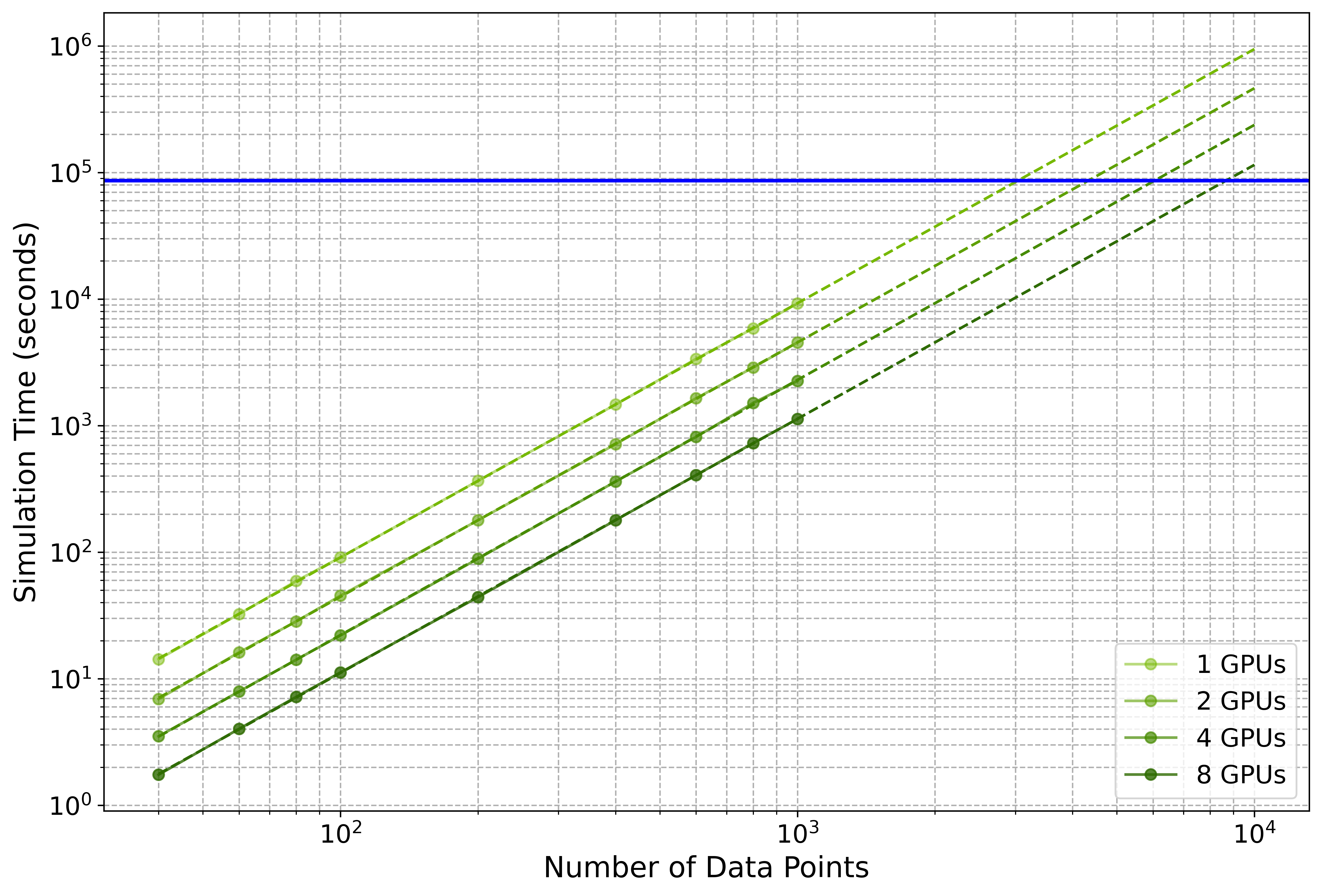}
    \caption{Execution time for quantum simulations against qubit count for a single A100 GPU and MPI-based 2, 4 and 8 multi-GPU setups. The performance enhancement with additional GPUs is evident, underscoring the benefits of parallelized computation.}
    \label{fig:multigpu-data}
\end{figure}

Figure \ref{fig:multigpu-data} presents a comparative analysis of computational time across different configurations, ranging from a single GPU (A100, 80GB) to 2, 4, and 8 multi-GPU arrangements using MPI for processing datasets of various sizes. The results distinctly highlight the superior efficiency and scalability of multi-GPU systems, especially when managing large-scale datasets. A notable reduction in processing time is observed with the integration of an 8-GPU setup, underscoring the considerable advantages of parallel computing for large-scale data analysis. In Figure \ref{fig:multigpu-data}, experimental data (solid line) from 40 to 1,000 data points is extrapolated to estimate the processing time for 10,000 data points, corresponding to nearly 50 million circuits (dashed line). The projection indicates that an eight-GPU system could achieve linear acceleration, reducing a week-long processing task using the simulated QSVM to approximately one day (blue line).

\section{Conclusion}
 This paper has presented a comprehensive investigation into the feasibility and performance of large-scale QSVM simulations using a tensor-network-based framework integrated with NVIDIA’s cuQuantum SDK. By leveraging the cuTensorNet library on multi-GPU platforms, we significantly reduced the otherwise prohibitive computational overhead associated with simulating large qubit systems. Rigorous performance benchmarks demonstrated not only near-quadratic scaling for circuit simulations—thereby overcoming the exponential barriers of conventional state-vector approaches—but also robust speedups via MPI-based parallelization for quantum circuit simulation. Moreover, our experiments on benchmark datasets, including MNIST and Fashion-MNIST, revealed that QSVMs can achieve high classification accuracy, emphasizing the promise of quantum methods for complex, high-dimensional data. Crucially, the observed improvements in accuracy with increasing dataset size underscore the value of scalable simulation environments as a test bed for algorithmic refinements and real-world applications. The successful integration of cuTensorNet and multi-GPU infrastructures thus serves as a critical validation of quantum–HPC synergy, pointing to a practical route toward bridging near-term quantum hardware limitations and large-scale quantum machine learning goals. These results lay a foundation for further advances in high-performance quantum simulations and reinforce the potential impact of quantum-enhanced algorithms within the rapidly evolving Quantum–HPC ecosystem.

\section*{Data Availability}
The code supporting the findings of this research is available on GitHub at the following repository: \url{https://github.com/Tim-Li/cuTN-QSVM}. This repository includes the scripts and data required to reproduce the results presented in this paper.

\section*{Conflict of Interest Statement}
The authors declare that they have no conflict of interest related to this work.

\section*{Acknowledgment}
The authors express their gratitude to W.C. Qian (MediaTek) for invaluable assistance and insights, which were pivotal to the success of this research. This research was funded by the U.K. Engineering and Physical Sciences Research Council under Grant No. EP/W032643/1. K.C. acknowledges financial support from the Turing Scheme for the Imperial Global Fellows Fund and the Taiwanese Government Scholarship to Study Abroad. This work was also supported by the National Science and Technology Council (NSTC), Taiwan, under Grants NSTC 112-2119-M-007-008- and NSTC 113-2119-M-007-013-. The authors thank the National Center for High-performance Computing of Taiwan for providing computational and storage resources, as well as the NVAITC and NVIDIA Quantum team for their technical support.

\section*{References}
\bibliographystyle{unsrt}  
\bibliography{mybib}

\end{document}